\documentclass[a4paper,10pt,twocolumn]{article} 
\usepackage{graphicx}
\usepackage{subfigure}
\usepackage{amsmath}
\usepackage{amssymb}
\usepackage{wrapfig}
\usepackage{color}
\usepackage[english]{babel}
\usepackage[warning,math]{easyeqn}
\usepackage[thinlines]{easybmat}
\begin{document}

\title{\textbf{Has the density of sources of gamma-ray burts\\ been constant over the last ten billion years ?}}

\author{Yves-Henri Sanejouand\footnote{yves-henri.sanejouand@univ-nantes.fr}\\ \\
        UMR 6286 of CNRS,\\
        Facult\'e des Sciences et des Techniques,\\
        Nantes, France.} 
\maketitle


\section*{Abstract}

A generic tired-light mechanism is examined in which a photon, like any particle moving in a medium, experiences friction, that is, a force resisting its motion. If the velocity of light is assumed to be constant, this hypothesis yields a Hubble-like law which is also a consequence of the $R_h = c t$ cosmology. Herein, it is used for estimating matter density as a function of redshift, allowing to show that the density of sources of long gamma-ray bursts appears to be nearly constant, up to $z \approx 4$. Assuming that the later is a fair probe of the former, this means that matter density has been roughly constant over the last ten billion years, implying that, at least over this period, matter has been in an overall state of equilibrium.

\vspace{0.5cm}
\noindent
Keywords: alternative cosmologies, tired-light model, matter density, long gamma-ray bursts.

\section*{Introduction}

The current flavour of the cosmological theory proposed by Georges Lemaitre \cite{Lemaitre:27}, namely, the $\Lambda$ cold dark matter paradigm ($\Lambda$CDM), is able to account for numerous observations, of various origins \cite{Ostriker:95,Tegmark:01}. However, $\Lambda$CDM relies on hypotheses about what the Universe is made of that may prove \textit{ad hoc} like, for instance, that $\approx \frac{3}{4}$ of the energy of the Universe is a so-called dark energy \cite{Frieman:08,Bartelmann:10,Planck:14} of unknown nature, with properties as weird as a negative pressure \cite{Perlmutter:99b}, $\approx$ 90\% of the rest being in the form of a non-baryonic so-called dark matter that has managed to escape detection on Earth up to now \cite{Bergstrom:00,Sigl:15}. Looking for alternative cosmologies \cite{Pasquier:01,Odintsov:07,Lima:10} may thus prove fruitful.

A central tenet of the family of cosmological theories initiated by Lemaitre is the postulated property that the wavelength of a photon expands as the scaling factor does. This postulate actually allowed Lemaitre to anticipate the discovery of the redshift-distance law \cite{Lemaitre:27} by Edwin Hubble \cite{Hubble:29}. It noteworthy implies that, within the frame of Lemaitre cosmologies, matter density is a straightforward function of redshift. 

Hereafter, it is assumed that this postulate is actually wrong, namely, that the Hubble law has a physical cause of a different nature.

As alternative explanations of this law, a number of tired-light mechanisms have been proposed \cite{Finlay:54,Born:54,deBroglie:66,Vigier:88,North,Marmet:18}, the first one by Fritz Zwicky, in 1929 \cite{Zwicky:29}. Herein, a generic model is considered. 

In such a context, there is no reason to assume that matter density varies with the same trend as in Lemaitre cosmologies. Reciprocally, 
providing estimates for this quantity as a function of redshift should allow to gain key insights about the underlying cosmology. 
To do so, it is necessary to examine a category of representative objects whose redshift is known over a wide range, with few selection biases. In both respects, thanks to the \textit{Swift} mission \cite{Swift}, sources of gamma-ray bursts (GRBs) are attractive candidates.          

\section*{The tired-light model}

Let us assume that a photon, like any particle moving in a medium, experiences friction, that is, a force, $f$, resisting its motion, so that:   
\begin{equation}
f = -m_0 \beta_H c_0
\label{eq:friction}
\end{equation}
where $m_0$ is the mass carried by the photon, $\beta_H$, its damping constant, $c_0$ being the speed of light. If $f$ is assumed to be constant over time, then the photon looses energy during its travel towards an observer in such a way that:
\begin{equation}
h \nu_{obs} = h \nu_0 + f D_c
\label{eq:energy}
\end{equation}
where $\nu_{obs}$ is the frequency of the photon measured by the observer, $\nu_0$, its frequency when it is emitted, $D_c$, the distance between the source of the photon and the observer, $h$ being the Planck constant. Then, with \cite{Einstein:05mc}:
\begin{equation}
m_0 = \frac{h \nu_0}{c_0^2}
\label{eq:ldb}
\end{equation}
eqn (\ref{eq:energy}) yields, together with eqn (\ref{eq:friction}) and (\ref{eq:ldb}):
\begin{equation}
\frac{\nu_0 - \nu_{obs}}{\nu_0} = \frac{z}{1+z} = \frac{\beta_H}{c_0}  D_c
\label{eq:hubble}
\end{equation}
where $z$ is the observed redshift. When $z \ll 1$, eqn (\ref{eq:hubble}) can be approximated by:
$$
z \approx \frac{\beta_H}{c_0} D_c
$$
that is, a relationship of the same form as the one discovered by Hubble~\cite{Hubble:29}. So, let us assume that the cosmological redshift is mostly due to photon friction, as defined above, namely, that:
\begin{equation}
\beta_H = H_0
\label{eq:hyp}
\end{equation} 
where $H_0$ is the Hubble constant. 

Then, $\rho_M(z_i,z_j)$, the matter density in the redshift range $\Delta z = z_j - z_i$, can be obtained from counts of representative objects as follows:
\begin{equation}
\rho_M(z_i,z_j) = \frac{n(z_i,z_j)}{V(z_j) - V (z_i)}
\label{eq:density}
\end{equation}
where $n(z_i,z_j)$ is the number of objects with a redshift between $z_i$ and $z_j$, $V(z_j)$ being the volume of the sphere including objects with $z \leq z_j$.

If the speed of the light is assumed to be constant, eqn (\ref{eq:hubble}-\ref{eq:density}) yield: 
\begin{equation}
\rho_M(z_i,z_j) = \frac{n(z_i,z_j)}{\frac{4}{3} \pi D_H^3} ( \frac{1}{ \frac{z_j^3}{(1+z_j)^3} - \frac{z_i^3}{(1+z_i)^3}} )
\label{eq:dens}
\end{equation}
where $D_H=\frac{c_0}{H_0}$ is the Hubble length.

\section*{Data}

The 328 GRB sources whose redshift has been determined by \textit{Swift} \cite{Derek:09} with fair accuracy\footnote{On april 2017 (http://swift.gsfc.nasa.gov).} were considered for the present analysis. Since long and short GRB ($t_{90} < 0.8 s$) are expected to have different physical origins \cite{Sari:13}, the 27 later ones were disregarded.
GRB 040923 was also put aside. Being at $z = 8.23$ \cite{Davies:14}, that is, significantly farer than other GRBs in our dataset, it may indeed prove atypical.

\begin{figure}[t]
  \includegraphics[width=8.0 cm]{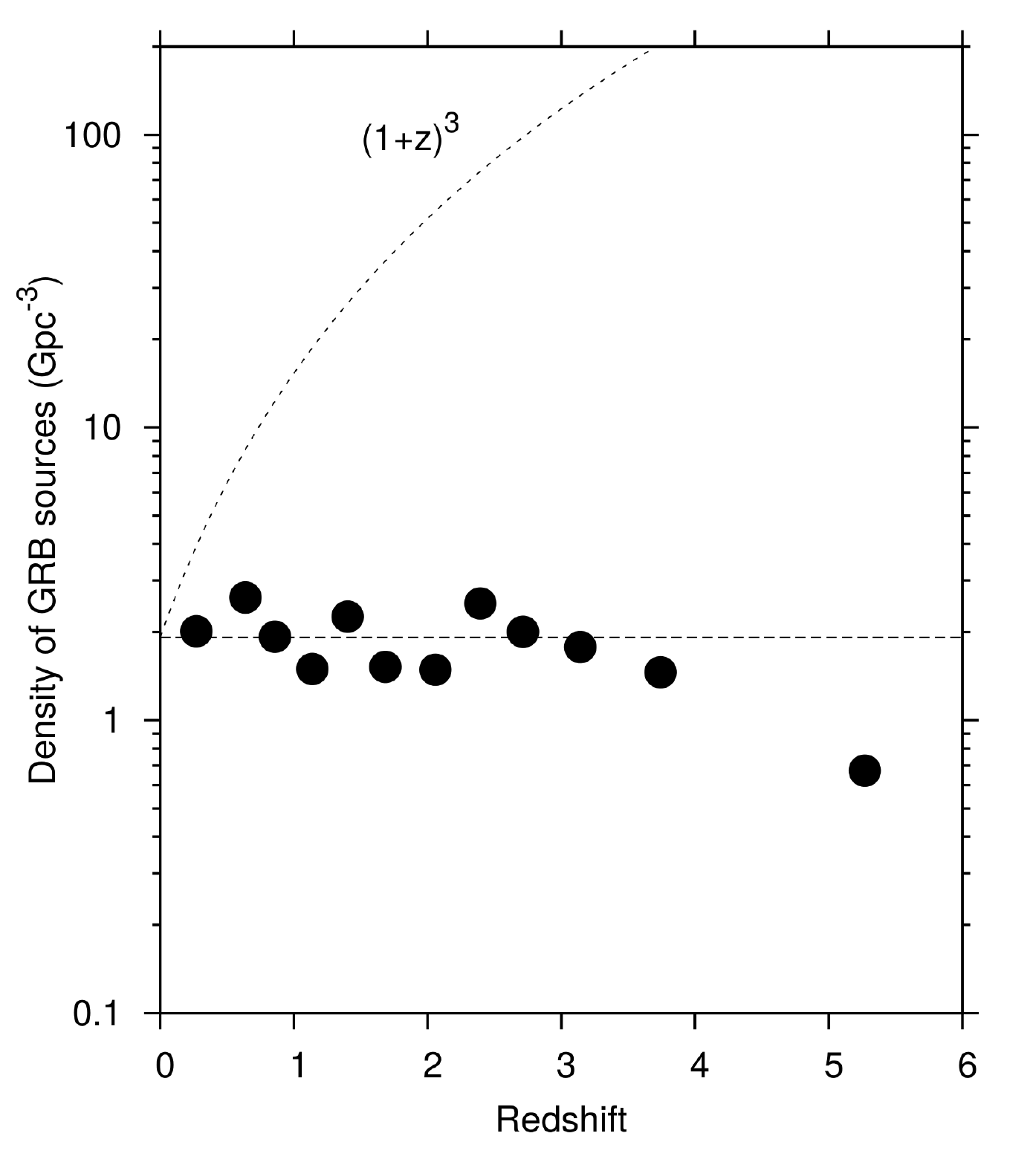}
   \caption{Density of sources of long gamma-ray bursts observed by \textit{Swift}, as a function of redshift. 
For a given range of redshift, the density (filled circles) is obtained through the volume occupied by 25 GRB sources, according to the tired-light model advocated herein. Horizontal line: fit of these data for $z < 4$, when density is assumed to be constant. Dotted line: density expected if GRB sources are a fair probe of matter density, according to Lemaitre cosmologies.}
\label{Fig:grb}
\end{figure}

\section*{Results}

The redshifts of the 300 GRBs considered were sorted by increasing values, so as to build 12 groups of 25 GRBs, the GRBs of each group having redshifts belonging to a given range, namely: 
\begin{flalign*}
z \in &\left] 0, 0.54 \right], \left[ 0.54, 0.74 \right], 
       \left] 0.74, 0.98 \right], \left] 0.98, 1.29 \right],&\\
      &\left] 1.29, 1.51 \right], \left] 1.51, 1.86 \right], 
       \left] 1.86, 2.26 \right], \left] 2.26, 2.53 \right],&\\ 
      &\left] 2.53, 2.9 \right], \left] 2.9, 3.39 \right], 
       \left] 3.39, 4.1 \right], \left( 4.1, 6.44 \right]&
\end{flalign*}
Figure \ref{Fig:grb} shows the density of GRB sources observed by \textit{Swift} for each of these 12 ranges, as obtained using eqn ({\ref{eq:dens}). It is found to be fairly constant up to $z \approx 4$, with $\rho_M(z_i,z_j) = 1.8 \pm 0.5$ GRB sources per Gpc$^3$, if $H_0 =$ 73 km.s$^{-1}$.Mpc$^{-1}$ \cite{Riess:16,Paturel:17}.
 
Note that the decrease of density observed for $z > 4$ could be due to the sensitivity limits of \textit{Swift} or to the difficulty of redshift determination at such distances. Density fluctuations may also prove meaningful \cite{Sanejouand:14}. 

Remind that, within the frame of Lemaitre cosmologies, in sharp contrast with results shown in Figure \ref{Fig:grb}, matter density is expected to be 125 times larger at $z=4$ than at $z=0$. 

\section*{Discussion}

\subsection*{What is constant ?}

Figure \ref{Fig:grb} shows that, if Hubble's law is due to the tired-light mechanism advocated herein, then the density of GRB sources has been on average constant over the last ten billion years. Since long GRBs seem to be produced by a small subset of supernovae of type Ic \cite{Savchenko:15}, and since supernovae rates may prove proportional to host galaxy masses \cite{Modjaz:15}, this could mean that the total galactic mass has been nearly constant over this period. Assuming that the total galactic mass is a constant enough fraction of the whole matter content implies that matter density has been roughly constant as well.  

\subsection*{Are tired-light models still relevant ?}

It has been claimed that, as predicted by Lemaitre cosmologies \cite{Schrodinger:39nat,Schrodinger:39,Wilson:39}, the light curves of Type Ia supernovae are dilated by $( 1 + z )$ \cite{Riess:96,Perlmutter:05,Blondin:08} and that, as a consequence, theories in which photons dissipate their energy during travel are excluded \cite{Riess:96,Perlmutter:05}. However, no such time dilation was detected in GRB \cite{Petrosian:13} or quasar \cite{Hawkins:01,Hawkins:10} light curves, casting doubts on the generality of the phenomenon. 

\subsection*{An independent check of the model}

With $D_c = c_0 \Delta t$, eqn (\ref{eq:hubble}) and (\ref{eq:hyp}) yield:
$$
\frac{z}{1+z} = H_0 \Delta t
$$
where $\Delta t$ is the photon time-of-flight between the source and the observer. This relationship noteworthy implies that \cite{Sanejouand:14}:
\begin{equation}
H(z)=-\frac{1}{1+z} \frac{\partial z}{\partial t}=H_0 (1+z)
\label{eq:hz}
\end{equation}
This prediction of the model considered herein is actually shared by the $R_h = c t$ cosmology developed by Fulvio Melia and his collaborators \cite{Melia:12} who have claimed that, compared to the $\Lambda$CDM prediction, it is favoured by various criteria \cite{Melia:13,Melia:18}. 
As a matter of fact, at z=1.965, it has revently been found that $H(z)=$ 186.5 $\pm$ 50.4 \cite{Moresco:15}, while eqn (\ref{eq:hz}) predicts  $H(z)=$ 216.5 $\pm$ 5.3, if $H_0 =$ 73.02 $\pm$ 1.79 km.s$^{-1}$.Mpc$^{-1}$ \cite{Riess:16}.

\section*{Conclusions}

When the Hubble law is explained through a generic tired-light mechanism, the density of GRB sources is found to be nearly constant (Figure \ref{Fig:grb}), up to z $\approx$ 4 at least.  
This means that matter density may have been roughly constant over the last ten billion years, implying that, at least over this period, matter has been in an overall state of equilibrium. In turn, this would mean that, at very large distances, gravitation is counterbalanced by some yet unknown physical phenomenon.      

Remind that such a line of thought has already been put forward a number of times \cite{North}, based on various grounds \cite{Einstein:17,Bondi:48,Hoyle:48}.



\end{document}